\documentclass[lettersize,journal]{IEEEtran}
\usepackage{amsmath,amsfonts}
\usepackage{algorithmic}
\usepackage{algorithm}
\usepackage{array}
\usepackage{cuted}
\usepackage[caption=false,font=normalsize,labelfont=sf,textfont=sf]{subfig}
\usepackage{textcomp}
\usepackage{stfloats}
\usepackage{url}
\usepackage{verbatim}
\usepackage{graphicx}
\usepackage{cite}
\usepackage{amssymb}
\usepackage{graphicx}
\usepackage{algorithm}
\usepackage{algorithmic}
\usepackage{cases}
\usepackage{diagbox}
\usepackage{epstopdf}
\usepackage{geometry, tabularx}
\usepackage{multirow}
\usepackage{enumerate}
\usepackage{bm}
\usepackage{pifont}
\usepackage{footnote}
\usepackage{float}
\usepackage{titlesec}
\usepackage{setspace}
\usepackage{makecell}
\usepackage{mathrsfs}
\usepackage[utf8]{inputenc} % allow utf-8 input
\usepackage[T1]{fontenc}    % use 8-bit T1 fonts
\usepackage{hyperref}       % hyperlinks
\usepackage{url}            % simple URL typesetting
\usepackage{booktabs}       % professional-quality tables
\usepackage{amsfonts}       % blackboard math symbols
\usepackage{nicefrac}       % compact symbols for 1/2, etc.
\usepackage{microtype}      % microtypography
\usepackage{xcolor}         % colors

\newtheorem{thm}{\bf Theorem}

\begin{document}

\title{A Partition-insensitive Parallel Framework for Distributed Model Fitting}

\author{Xiaofei Wu\thanks{X. Wu is with the College of Mathematics and Statistics, Chongqing University, Chongqing 401331, China (e-mail: xfwu1016@163.com)},
 Rongmei Liang\thanks{R. Liang is with the College of Mathematical Medicine, Zhejiang Normal University, Jinhua 321004, China (e-mail: liang\_r\_m@163.com)},
 Fabio Roli\thanks{F. Roli is with the University of Genova, Genoa 16167, Italy (e-mail: fabio.roli@unige.it)},
 Marcello Pelillo\thanks{M. Pelillo is with the University of Venice,Venice 30176, Italy (e-mail: pelillo@unive.it)}, 
and Jing Yuan\thanks{J. Yuan is with the College of Mathematical Medicine, Zhejiang Normal University, Jinhua 321004, China (e-mail: jyuan@zjnu.edu.cn)}}

%~\IEEEmembership{Fellow,~IEEE,}

        % <-this % stops a space
%\thanks{This paper was produced by the IEEE Publication Technology Group. They are in Piscataway, NJ.}% <-this % stops a space
%\thanks{Manuscript received April 19, 2021; revised August 16, 2021.}}正式投稿时这两行就不要注释了。

% The paper headers(正式投稿的时候下面的内容就不要注释了)


\maketitle

\begin{abstract}
Distributed model fitting refers to the process of fitting a mathematical or statistical model to the data using distributed computing resources, such that computing tasks are divided among multiple interconnected computers or nodes, often organized in a cluster or network. Most of the existing methods for distributed model fitting are to formulate it in a consensus optimization problem, and then build up algorithms based on the alternating direction method of multipliers (ADMM). This paper introduces a novel parallel framework for achieving a distributed model fitting. In contrast to previous consensus frameworks, the introduced parallel framework offers two notable advantages. Firstly, it exhibits insensitivity to sample partitioning, meaning that the solution of the algorithm remains unaffected by variations in the number of slave nodes or/and the amount of data each node carries. Secondly, fewer variables are required to be updated at each iteration, so that the proposed parallel framework performs in a more succinct and efficient way, and  adapts to high-dimensional data. In addition, we prove that the algorithms under the new parallel framework have a worst-case linear convergence rate in theory. Numerical experiments confirm the generality, robustness, and accuracy of our proposed parallel framework.
\end{abstract}


\section{Introduction}\label{sec1}

A general model fitting problem can be expressed as follows:
\begin{align}\label{in1}
\min_{\bm{x}}{\mathcal{L}(\bm{Ax},  \bm{b}) + \mathcal{R}(\bm{x})},
\end{align}
where $\bm{x} \in \mathbb{R}^n$ is the coefficient to be estimated, $\bm{A} \in \mathbb{R}^{m \times n}$ represents the feature matrix, $\bm{b} \in \mathbb{R}^m$ denotes the output vector, $\mathcal{L}: \mathbb{R}^m \rightarrow \mathbb{R}$ stands for the loss function, and $\mathcal{R}$ denotes the regularization function.
Typically, $\mathcal{L}$ is assumed to be additive, implying
\begin{align}\label{in2}
\mathcal{L}(\bm{Ax}, \bm{b}) = \sum_{i=1}^m \mathcal{L}_i(\bm{a}_i^\top \bm{x}, b_i),
\end{align}
where $\mathcal{L}_i: \mathbb{R} \rightarrow \mathbb{R}$ represents the loss for the $i$-th training example, $\bm{a}_i \in \mathbb{R}^n$ and $b_i \in \mathbb{R}$ respectively denote the feature vector and the output or response for example $i$. Each $\mathcal{L}_i$ can vary, although in practice, they are usually uniform.
We also assume that the regularization function $\mathcal{R}$ is separable or partially separable. 
The most common examples of regularization in statistical settings include Tikhonov regularization, often referred to as a ridge penalty, expressed as $\mathcal{R}(\bm x) = \lambda \lVert \bm x \rVert_2^2$ \cite{Hoerl1970Ridge}; the $\mathcal{R} = \lambda \lVert \bm x \rVert_1$ penalty, commonly known as the lasso penalty \cite{Tibshirani1996Regression}; and group regularization represented by $\mathcal{R}(\bm x) = \lambda \lVert \bm x \rVert_{2,1} = \lambda \sum_{g=1}^{G} \left\| \bm x_g \right\|_{2}$, where $\bm x$ is partitioned into $G$ disjoint groups \cite{Yuan2006Model}. Obviously, when $G=n$, $\lVert \bm x \rVert_{2,1}$ is $\lVert \bm x \rVert_1$. Here, $\lambda$ denotes a positive regularization parameter. Regularization parameter can also be more  elaborate. For instance,  $\bm \lambda$ could be represented as a weighted vector, introducing a weighted form of the regularization term. This results in expressions such as $\mathcal{R}(\bm{x}) = \lVert \bm{\lambda} \odot \bm{x} \rVert_2^2$, $\lVert \bm{\lambda} \odot \bm{x} \rVert_1$ and  $\sum_{g=1}^{G} \lambda_g \left\| \bm x_g \right\|_{2}$, where $\odot$ denotes the Hadamard product. As described by \cite{Boyd2010Distributed},  the model fitting problem in (\ref{in1}) can encompass many practical classification and regression models.

In parallel computing environments, the data is assumed to be divided into the following ways
\begin{align}\label{data}
\bm A = [\bm A_1^\top, \dots,\bm A_D^\top  ]^\top \ \text{and} \ \bm b = (\bm b_1,\dots, \bm b_D)^\top,
\end{align}
with $\bm A_d \in  \mathbb{R}^{m_d \times n}$ and  $\bm b_d \in  \mathbb{R}^{m_d}$, where $\sum_{d=1}^D m_d = m$.
The distributed storage of data,  as well as the non-smoothness of both $\mathcal{L}$ and $\mathcal{R}$,  pose some challenges in designing algorithms to solve  problem (\ref{in1}). \cite{Boyd2010Distributed}
 proposed a consensus-based ADMM algorithm to facilitate model fitting  in a parallel manner. Each node typically handles a subset of the data or performs specific computations, and the results are subsequently aggregated to derive the final model parameters or estimates. This approach proves advantageous in situations where the volume of training examples is substantial, rendering it impractical or unfeasible to process them on a single machine. Moreover, it finds utility in scenarios where data is inherently distributed, such as in recommender systems \cite{Zhang2022DS}, online social network data \cite{Frana2020Distributed},  wireless sensor networks \cite{Yuan2015Communication}, and various cloud computing applications \cite{Feng2017An}.

However, parallel ADMM and its variants for solving the consensus problem encounter an undeniable drawback: excessive auxiliary variables during the iteration process, especially when dealing with  high-dimensional dataset (large $m$ and $n>m$). Although these auxiliary variables make solving some subproblems easier, introducing excessive variables can cause slow convergence of the algorithm \cite{Lin2022Alternating}, and may affect the accuracy of the solution \cite{Fan2021Penalized}. The reason for the excessive auxiliary variables is due to the parallel structure of consensus.
To be specific, for (\ref{in1}), its consensus optimization problem is as follows,
\begin{align}\label{in4}
\min_{\bm{x},\{\bm{z}_d\}} {{\kern 1pt} {\kern 1pt} {\kern 1pt} {\kern 1pt} \sum_{d=1}^D \mathcal{L}(\bm{A}_d \bm{z}_d, \bm{b}_d) +  \mathcal{R}(\bm{x})}\\
\text{s.t.}{\kern 1pt} {\kern 1pt} {\kern 1pt} {\kern 1pt} {\kern 1pt} {\kern 1pt} \bm{x} = \bm{z}_d, d=1,2,\dots,D, \notag
\end{align}
where $ \{ \bm x = \bm{z}_d\}$  is the consensus constraint added for parallel structures.  Note that our original optimization problem (\ref{in1}) only needs to obtain the solution $\bm x$ that is a vector with dimension $n$.  On the other hand, for consensus optimization problems in (\ref{in4}),  to achieve parallel structures, the total dimensions of the solutions that need to be solved are $(D+1) n$ dimensions. In addition to this common parallel structure, many other centralized and decentralized parallel structures have been proposed to enhance the performance of consensus-based ADMM algorithms and the communication efficiency of nodes, as presented in works such as  \cite{Yan2020Parallel}, \cite{Lin2022Alternating} and \cite{Qiu2023PSRA}.  However, these improved parallel structures still cannot solve the problem of excessive optimization variables, and even in decentralized structures, this problem is further exacerbated.

In this paper, we propose a new parallel framework to reduce the problem of excessive auxiliary variables in existing consensus parallel ADMM algorithms. The main idea of the new algorithm framework is to remove these redundant auxiliary variables and simply achieve algorithm parallelism from the perspective of variable splitting. This consideration is reasonable due to the fact that  for distributed model fitting,   we only want to obtain $\bm x$, and $\{\bm z_d\}$ is only for the purpose of implementing parallel structures.  The proposed new parallel framework not only avoids adding too many auxiliary variables, but also has a surprising property, which is the insensitivity of sample partitioning. We will theoretically prove that no matter how the feature matrix $\bm A$ is divided by rows, the convergence solution of  parallel algorithm  designed according to our framework  changes very little, or even no change. This characteristic ensures the stability of the convergence solution, even when confronted with numerous parallel machines.

The rest of the paper is organized as follows. Section \ref{sec2}  introduces some proximal  operators and consensus based ADMM algorithm. In Section \ref{sec3}, we propose our partition-insensitive parallel (PIP) algorithm framework. The theoretical guarantee of the algorithm framework is presented  in Section \ref{sec4}. Section \ref{sec7} summarizes the findings and concludes the paper, with a discussion on future research directions. This paper also provides some application scenarios of algorithm frameworks in Section \ref{sec5}, such as regularized logistic regression, regularized SVMs, and regularized linear regression models. And the Numerical experiments are provided in Section \ref{sec6}.

\section{Preliminaries}\label{sec2}
\subsection{Proximal operator}\label{sec2.1}
In iterative algorithms, the existence of closed-form solutions for the proximal operator has a significant impact on the algorithm efficiency. 
To facilitate the closed-form solutions for the updates in the ADMM algorithm, we introduce the following proximal operators,
\begin{align}\label{po}
\operatorname{prox}_{\gamma, g}(\bm a) = \arg \min_{\bm c} \left\{ g(\bm c) + \frac{\gamma}{2}\|\bm c - \bm a\|_2^2\right \},
\end{align}
where $g(\bm c)$ is a  nonnegative real-valued function,  $\bm a$ is a  constant vector, and $\mu > 0$ is a constant. In this paper, the function $g(\bm c)$ is always separable or additive, meaning $g(\bm c) = \sum_{j}g(c_j)$. Then, we can  divide the optimization problem (\ref{po}) into multiple independent univariate problems.

Many widely used operators in applications are actually special cases of proximal operators, among which the most famous one is the soft-thresholding operator in \cite{Donoho1995De}. The soft-thresholding operator is defined as
$$\text{prox}_{\gamma,  \lambda \| \bm c \|_{1}} (\bm a) = \arg \min_{c} \left( \lambda\| \bm c \|_{1} + \frac{\gamma}{2} \| \bm c - \bm a \|_2^2 \right).$$
The closed-form solution for $\text{prox}_{\gamma, \lambda \| \bm c \|_{1}} (\bm a)$ is given by
\begin{align}\label{lasso}
\text{prox}_{\gamma,  \lambda \| \bm c \|_{1}} (\bm a) = \text{sign}(\bm a) \odot \max(|\bm a| - \frac{\lambda}{\gamma}, 0),
\end{align}
where sign$(\cdot)$ function is defined component-wise such that sign$(t) = 1$ if $t > 0$, sign$(t) = 0$ if $t = 0$, and sign$(t) = -1$ if $t < 0$.

Next, we need to consider the proximal operator commonly used for $\ell_{2,1}$-norm regularization, which is also known as the group soft-thresholding operator  in \cite{Boyd2010Distributed}. It is defined as $$\text{prox}_{\gamma,\lambda \| \bm c \|_{2,1}}(\bm{a}) = \arg \min_{\bm c} \left( \lambda\| \bm c \|_{2,1} + \frac{\gamma}{2} \| \bm c - \bm a \|_2^2 \right).$$
The closed-form solution for $\text{prox}_{\gamma,\lambda \| \bm c \|_{2,1}}(\bm{a})$ is given by
\begin{align}\label{gglasso}
\text{prox}_{\gamma,\lambda \| \bm c \|_{2,1}}(\bm{a}) = \frac{\bm a} {\|\bm a \|_2} \cdot \max(\|\bm a \|_2- \frac{\lambda}{\gamma} ,0).
\end{align}
When each group has only one element, (\ref{gglasso}) is (\ref{lasso}).

Finally, the closed-form solution for the proximal operator of $\ell_{2}$-norm regularization ($\lambda \| \bm c \|_{2}^2$) is readily derived thanks to the quadratic differentiability of $\ell_{2}$. Specifically, it is given by
\begin{align}\label{l2}
\text{prox}_{\gamma,\lambda \| \bm c \|_{2}}(\bm{a}) = \frac{\gamma \bm a} {2\lambda +\gamma}. 
\end{align}

\subsection{Consensus parallel  ADMM}\label{sec2.2}

\begin{figure*}[h]\centering
\rotatebox{90}{\includegraphics[scale=0.6]{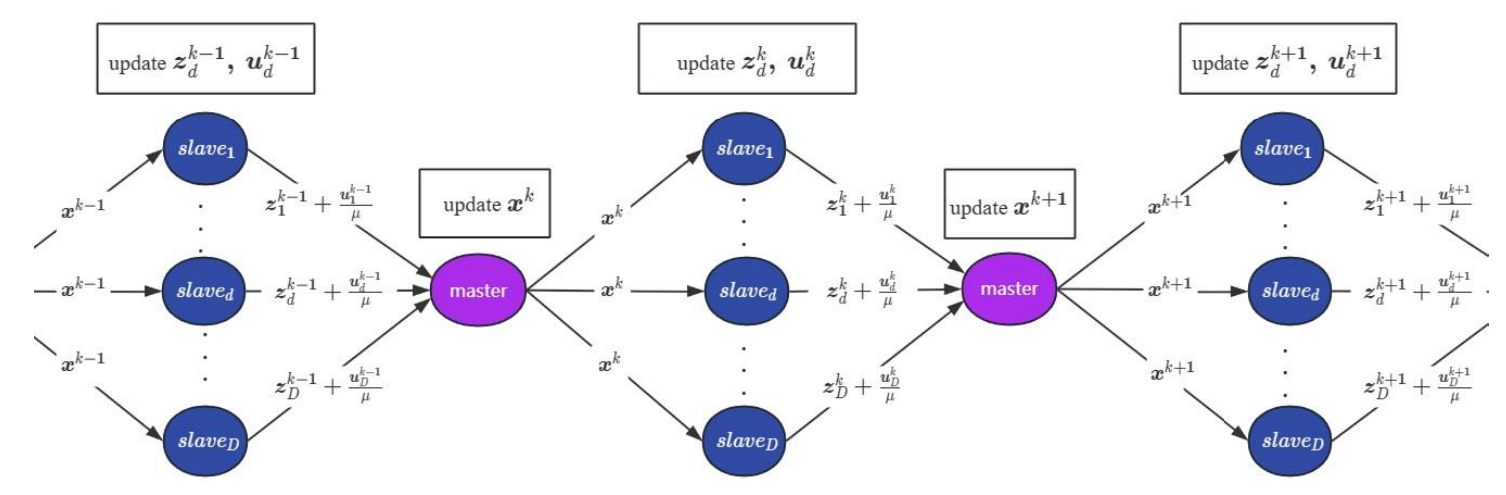}}
	\caption{Schematic diagram of master-slave node operation implemented in consensus based ADMM algorithm. The master node needs to receive $\bm z_d + \bm u_d/\mu$ posted by each slave node, while each slave node needs to receive $\bm x$ from the master node.}
    \label{fig1}
\end{figure*}

In this section, we will briefly introduce how existing consensus parallel  ADMM algorithms  in \cite{Boyd2010Distributed} can achieve distributed model fitting. And the schematic diagram of the implementation of the consensus parallel ADMM algorithm is shown in Figure \ref{fig1}.

The augmented Lagrangian of (\ref{in4}) is
\begin{align}\label{sec2eq1}
\begin{array}{l}
L_{\mu}(\bm x, \{\bm{z}_d\},\{\bm{u}_d\}) =  \sum_{d=1}^D \mathcal{L}(\bm{A}_d \bm{z}_d, \bm{b}_d) +  \mathcal{R}(\bm{x}) \\
- \sum_{d=1}^D\bm{u}_d^\top(\bm{x} - \bm{z}_d) + \frac{\mu}{2}\sum_{d=1}^D \left\| \bm{x} - \bm{z}_d \right\|_2^2,
\end{array}
\end{align}
where $\{\bm u_d\} \in \mathbb R^n$ represent the dual variables, and $\mu > 0$ is a tunable augmentation parameter. With  given $\{\bm z_d^{0}\}$ and $\{\bm u_d^{0}\}$,  the iterative scheme of parallel ADMM for (\ref{sec2eq1}) is as follow, where $d = 1,\dots,D$.
\begin{equation}\label{sec2eq2}
\left\{ \begin{array}{l}
\bm x^{k+1} \ =\mathop {\arg \min }\limits_{\bm x} \left\{ L_\mu(\bm x, \{\bm{z}_d^{k}\},\{\bm{u}_d^{k}\}) \right\},\\
\bm z_d^{k+1} \ =\mathop {\arg \min }\limits_{\bm z_d} \left\{ L_\mu(\bm x^{k+1}, \{\bm{z}_d\},\{\bm{u}_d^{k}\})  \right\},\\
\bm{u}_d^{k+1}=\bm{u}_d^{k}-\mu(\bm{x}^{k+1} - \bm{z}_d^{k+1}).
\end{array} \right.
\end{equation}
$\bm x^{k+1}$ is updated by the master node, while $\bm z_d^{k+1}$ and $\bm{u}_d^{k+1}$ are updated in parallel at each slave node. Note that updating dual variable $\bm u_d^{k+1}$ is a linear algebraic operation, it follows that the efficacy of parallel ADMM in (\ref{sec2eq2}) depends on the efficiency with which the first and second subproblems can be resolved.

$\bullet$ \textbf{Updating $\bm x^{k+1}$}. Receive $\{\bm{z}^k_d\}$ and $\{\bm{u}^k_d\}$ passed by each slave node. For the first subproblem in (\ref{sec2eq2}), it can be simplified as
\begin{align}\label{xupdate0}
\bm x^{k+1} \ =\mathop {\arg \min }\limits_{\bm x} \left\{ \mathcal{R}(\bm{x}) + \frac{\mu}{2} \left\| \bm{x} - \Bar{\bm{z}}^k_d  - \Bar{\bm{u}}^k_d/\mu  \right\|_2^2 \right\},
\end{align}
where $\Bar{\bm{z}}^k_d = \frac{1}{D}\sum_{d=1}^D \bm{z}^k_d $ and $\Bar{\bm{u}}^k_d = \frac{1}{D}\sum_{d=1}^D \bm{u}^k_d $.

$\bullet$ \textbf{Updating $\bm{z}_d^{k+1}$}. Receive $\bm{x}^{k+1}$  passed by the master node. For the  second subproblem in (\ref{sec2eq2}), we update it in parallel as 
\begin{equation}\label{cz}
\begin{aligned}
\bm z_d^{k+1} = & \mathop {\arg \min }\limits_{\bm z_d} \{ \mathcal{L}(\bm{A}_d \bm{z}_d, \bm{b}_d)\\  
&+ \frac{\mu}{2} \left\| \bm{x}^{k+1} - {\bm{z}}_d  - {\bm{u}}^k_d/\mu  \right\|_2^2  \}. 
\end{aligned}
\end{equation}

It is not difficult to see that the subproblem of $\bm x^{k+1}$ is a proximal operator. When $\mathcal{R}(\bm{x})$ corresponds to ridge ($\lVert \bm{x} \rVert_2^2$), Lasso ($\lVert \bm{x} \rVert_1$), or group penalty ($\lVert \bm{x} \rVert_{2,1}$), the update for $\bm{x}^{k+1}$ has a closed-form solution, as discussed in Section \ref{sec2.1}.
As described by \cite{Boyd2010Distributed}, the iteration of $\bm z_d^{k+1}$ determines which type of model fitting objective the algorithm achieves. In other word, 
it can be used to achieve different model fitting  by changing the form of $\mathcal{L}(\bm{A}_d \bm{z}_d, \bm{b}_d)$. 
 In their example presented in Section 8.2, the sparse least squares regression is represented by $\mathcal{L}(\bm{A}_d \bm{z}_d, \bm{b}_d) = \frac{1}{2}(\bm{b}_d - \bm{A}_d \bm{z}_d)^2$, while sparse logistic regression is denoted as $\mathcal{L}(\bm{A}_d \bm{z}_d, \bm{b}_d) = \sum_{i=1}^{m_d}\log(1+\exp^{-\bm {b}_d[i] {\bm A_d}[i,]^\top \bm z_d })$, where $\bm {b}_d[i]$ denotes the $i$-th component of vector $\bm {b}_d$, and ${\bm A_d}[i,]$ represents the $i$-th row of the matrix ${\bm A_d}$. Similarly, sparse support vector machine (SVM) is formulated as $\mathcal{L}(\bm{A}_d \bm{z}_d, \bm{b}_d) = \sum_{i=1}^{m_d} (1 - \bm {b}_d[i] {\bm A_d}[i,]^\top \bm z_d )_+$.

Summarizing the variables that need to be iterated in consensus parallel ADMM, we have a total of $(2D+1)n$ variables. With a large number of features $n$, the implementation of the parallel algorithm will lead to significant memory and computational burden. In addition, as \cite{Lin2022Alternating} pointed out, these too many variables can cause slow convergence speed, further increasing the computational burden mentioned above.  For sparse regression problems, \cite{Fan2021Penalized} indicated that in the parallel structure of consensus, as the number of parallel slave nodes increases, the sparsity of the regression coefficients will decrease. We will propose a new parallel framework in the next section that can effectively avoid these two issues.

\section{Proposed Parallel Framrwork}\label{sec3}
\begin{figure*}[h]\centering
\rotatebox{-90}{\includegraphics[scale=0.6]{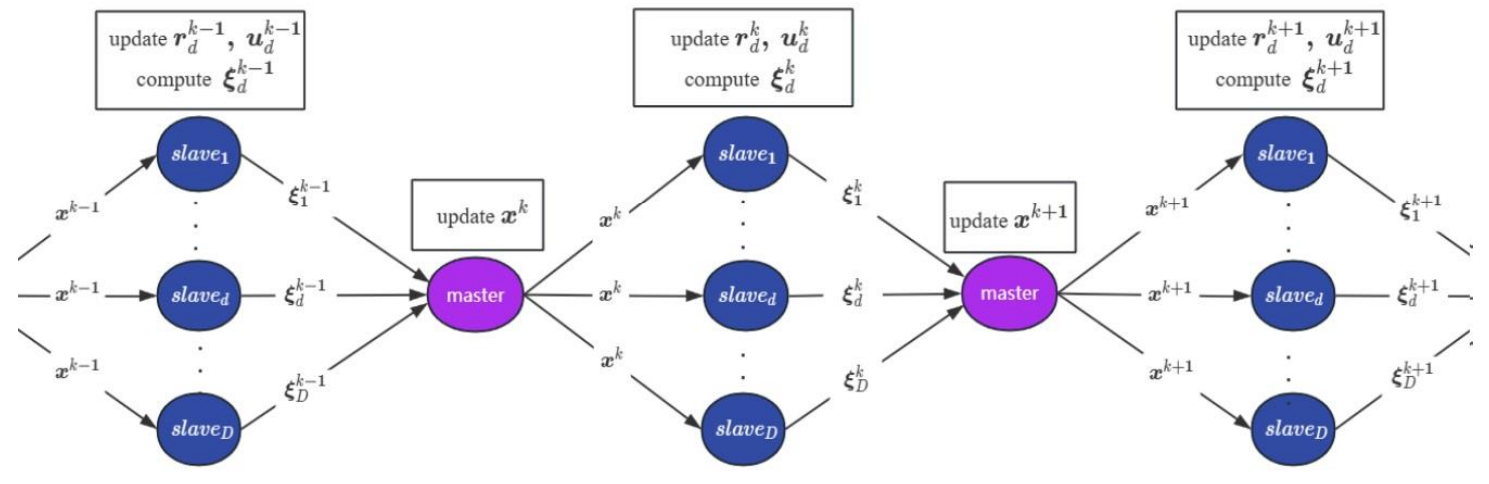}}
	\caption{Schematic diagram of master-slave node operation implemented in Algorithm \ref{alg1}. The master node needs to receive $\bm \xi_d$ posted by each slave node, while each slave node needs to receive $\bm x$ from the master node.}
    \label{fig2}
\end{figure*}
Looking back at the optimization form of the original problem in (\ref{in1}), if we do not add auxiliary variables, we can adopt the  method similar to \cite{Wu2023Partition}  and directly divide the samples to achieve parallelism.  The constraint optimization form of parallel algorithms is
\begin{equation}\label{sec3eq1}
\begin{aligned}
& \min_{\bm{x},\{\bm{r}_d\}} {{\kern 1pt} {\kern 1pt} {\kern 1pt} {\kern 1pt} \sum_{d=1}^D \mathcal{L}(\bm{r}_d,  \bm{b}_d) +  \mathcal{R}(\bm{x})}\\
& \text{s.t.}{\kern 1pt} {\kern 1pt} {\kern 1pt} {\kern 1pt} {\kern 1pt} {\kern 1pt} \bm{A}_d\bm x = \bm{r}_d, d=1,2,\dots,D. 
\end{aligned}
\end{equation}

The augmented Lagrangian of (\ref{sec3eq1}) is
\begin{align}\label{sec3eq2}
\begin{array}{l}
L_{\mu}(\bm x, \{\bm{r}_d\},\{\bm{u}_d\}) =  \sum_{d=1}^D \mathcal{L}(\bm{r}_d, \bm{b}_d) +  \mathcal{R}(\bm{x}) \\
- \sum_{d=1}^D\bm{u}_d^\top(\bm A_d\bm{x} - \bm{r}_d) + \frac{\mu}{2}\sum_{d=1}^D \left\| \bm A_d\bm{x} - \bm{r}_d \right\|_2^2.
\end{array}
\end{align}
 With  given $\{\bm r_d^{0}\}$ and $\{\bm u_d^{0}\}$,  the iterative scheme of the proposed parallel ADMM for (\ref{sec3eq1}) is as follow, where $d = 1,\dots,D.$
\begin{equation}\label{sec3eq3}
\left\{ \begin{array}{l}
\bm x^{k+1} \ =\mathop {\arg \min }\limits_{\bm x} \left\{ L_\mu(\bm x, \{\bm{r}_d^{k}\},\{\bm{u}_d^{k}\}) \right\},\\
\bm r_d^{k+1} \ =\mathop {\arg \min }\limits_{\bm r_d} \left\{ L_\mu(\bm x^{k+1}, \{\bm{r}_d\},\{\bm{u}_d^{k}\})  \right\},\\
\bm{u}_d^{k+1}=\bm{u}_d^{k}-\mu(\bm A_d\bm{x}^{k+1} - \bm{r}_d^{k+1}).
\end{array} \right.
\end{equation}
Although this iterative step can implement parallel structures like the ADMM algorithm introduced in Section \ref{sec2}, it may encounter some problems. We will describe them in the specific steps of the following iterations.

$\bullet$ \textbf{Updating $\bm x^{k+1}$}. Receive $\{\bm{r}^k_d\}$ and $\{\bm{u}^k_d\}$ passed by each slave node. For the first subproblem in (\ref{sec3eq3}), it can be simplified as
\begin{equation}\label{xupdate}
\begin{aligned}
\bm x^{k+1} \ =\mathop {\arg \min }\limits_{\bm x}\{& \mathcal{R}(\bm{x}) + \frac{\mu}{2} \sum_{d=1}^D \| \bm A_d \bm{x}\\
& - {\bm{r}}^k_d  - {\bm{u}}^k_d/\mu  \|_2^2 \}.
\end{aligned}
\end{equation}

$\bullet$ \textbf{Updating $\bm{r}_d^{k+1}$}. Receive $\bm{x}^{k+1}$  passed by the master node. For the  second subproblem in (\ref{sec3eq3}), we update it in parallel as 
\begin{equation}\label{rupdate}
\begin{aligned}
\bm r_d^{k+1}  =\mathop {\arg \min }\limits_{\bm r_d} \{& \mathcal{L}(\bm{r}_d, \bm{b}_d)  + \frac{\mu}{2}  \| \bm A_d \bm{x}^{k+1}\\
& - {\bm{r}}_d  - {\bm{u}}^k_d/\mu  \|_2^2 \}.
\end{aligned}
\end{equation}

Note that in the subproblem of $\bm x$ above, the master node requires data matrix $\bm A$, but obviously in the parallel structure of master slave, the master node does not need to load data. Consequently, this structure is deemed unsuitable for parallelism. In fact, due to this limitation, most existing parallel ADMM algorithms avoid adopting this parallel structure and instead opt for the consensus parallel structure introduced in Section \ref{sec2.2}.

\cite{Wu2023Partition} suggested using a linearization approach to avoid loading data when dealing with $\bm x$-subproblem. We will introduce this method in detail below to  achieve parallelism in the algorithm.

\subsection{The update of the first subproblem}\label{sec3.1}
The linearization technique has been widely applied in ADMM algorithms and has found applications in various fields \cite{Liang2024Linearized}. The fundamental idea behind the linearization technique is to linearize the nontrivial matrix quadratic term, thereby creating a proximal operator. 

Note that
\begin{align}\label{data2}
\bm r = (\bm r_1^\top, \dots,\bm r_D^\top )^\top \ \text{and} \ \bm u = (\bm u_1^\top,\dots, \bm u_D^\top)^\top,
\end{align}
it follows that 
\begin{align}
\left\| \bm A \bm{x} - {\bm{r}}  - {\bm{u}}/\mu  \right\|_2^2 =\sum_{d=1}^D \left\| \bm A_d \bm{x} - {\bm{r}}_d  - {\bm{u}}_d/\mu  \right\|_2^2.
\end{align}
Then, the update of $\bm x^{k+1}$ in (\ref{xupdate}) can be rewritten as
\begin{align}\label{xupdate1}
\bm x^{k+1} \ =\mathop {\arg \min }\limits_{\bm x} \left\{ \mathcal{R}(\bm{x}) + \frac{\mu}{2} \left\| \bm A \bm{x} - {\bm{r}^k}  - {\bm{u}^k}/\mu  \right\|_2^2 \right\}.
\end{align}
Obviously, $\bm A$ is the non trivial matrix that needs linearization. If $\eta$ is a constant that is larger than the maximum eigenvalue of matrix $\mu \bm A^\top \bm A$ ($\text{eigen}(\mu \bm A^\top \bm A )$), then the linearized subproblem $\bm x^{k+1}$ is

\begin{equation}\label{xupdate2}
\begin{aligned}
\bm x^{k+1} \ =\mathop {\arg \min }\limits_{\bm x} \{& \mathcal{R}(\bm{x}) + \frac{\mu}{2} \left\| \bm A \bm{x} - {\bm{r}^k}  - {\bm{u}^k}/\mu  \right\|_2^2 \\
&+  \frac{1}{2} \left\| \bm x - \bm x^k  \right\|_{\bm S}^2 \},
\end{aligned}
\end{equation}

where $\bm S = \eta \bm I_n  - \mu \bm A^\top \bm A$ and $\left\| \bm x - \bm x^k  \right\|_{\bm S}^2 = (\bm x - \bm x^k)^\top \bm S (\bm x - \bm x^k)$.
Omitting some constants unrelated to $\bm x$, (\ref{xupdate2}) can be reorganized as
\begin{equation}\label{xupdate3}
\begin{aligned}
\bm x^{k+1} \ =&\mathop {\arg \min }\limits_{\bm x} \{ \mathcal{R}(\bm{x}) + \frac{\eta}{2}\| \bm x - \bm x^k  \\ 
&+ \frac{\mu}{\eta}\bm A^\top (\bm A \bm x^k   - {\bm{r}^k}  - {\bm{u}^k}/\mu)  \|_2^2 \}.
\end{aligned}
\end{equation}

Let $\bm \xi_d^k =  \bm A_d^\top (\bm A_d \bm x^k  - {\bm{r}_d^k}  - {\bm{u}_d^k}/\mu)$.
Note also that for any $D$,
\begin{equation}\label{xixi}
\begin{aligned}
&\quad \bm A^\top (\bm A \bm x^k - {\bm{r}^k}  - {\bm{u}^k}/\mu) \\
&= \sum_{d=1}^D \bm A_d^\top (\bm A_d \bm x^k  - {\bm{r}_d^k}  - {\bm{u}_d^k}/\mu)\\
& = \sum_{d=1}^D \bm \xi_d^k,
\end{aligned}
\end{equation}
it follows that $\bm x^{k+1}$ can be obtained by solving 
\begin{align}\label{xxx}
\mathop {\arg \min }\limits_{\bm x} \left\{ \mathcal{R}(\bm{x}) + \frac{\eta}{2} \left\| \bm x - \bm x^k   + \frac{\mu}{\eta}\sum_{d=1}^D \bm \xi_d^k  \right\|_2^2 \right\}.
\end{align}
Therefore, the master node does not need to load the original data, and only needs to collect $\bm \xi_d^k $ from each slave node to complete the iteration of $\bm x^{k+1}$.
It is not difficult to see that (\ref{xxx}) is the same as (\ref{xupdate0}), except for some constant terms, and it is also a proximal operator with closed-form solution.

However, another issue arises: how to select the $\eta$ value. Let $\eta_d$ be the maximum eigenvalue of $\mu \bm A_d^\top \bm A_d$ ($\text{eigen}(\mu \bm A_d^\top \bm A_d )$). Due to the fact that $\sum_{d=1}^D \eta_d $ is large than $\text{eigen}(\mu \bm A^\top \bm A )$, as proposed by \cite{Wu2023Partition}, we can choose $\eta=\sum_{d=1}^D \eta_d$. In the algorithm's implementation, the power method proposed by \cite{Liang2024Linearized} can be utilized to compute $\eta_d$ on each slave node initially, which is then passed to the master node. During the iteration of the ADMM algorithm, $\eta_d$ only needs to be computed and passed once.

Although we drew inspiration from the algorithm designed by \cite{Wu2023Partition}, our algorithm  differs significantly from theirs. In \cite{Wu2023Partition}, the purpose of their paper is to solve regression problems, and the introduced auxiliary variable is residual. However, our paper focuses on model fitting problems, and the introduced constraint is $ \bm A \bm x= \bm r$. This makes our paper not only applicable to solving regression problems, but also capable of handling many classification problems, such as logistic regression, sparse SVMs, etc.

\subsection{The update of the second subproblem}\label{sec3.2}
For the $\bm r_d^{k+1}$-update in (\ref{rupdate}),  $\mathcal{L}(\bm{r}_d, \bm{b}_d)$ has different expressions for fitting  different models. For classification models, 
\begin{align*}
\mathcal{L}(\bm{r}_d, \bm{b}_d) = \sum_{i=1}^{m_d}\mathcal{L}(\bm{r}_d[i] \bm{b}_d[i]),
\end{align*}
with $\bm{b}_d \in \{-1,1\}^{m_d}$; 
and for regression models, 
 \begin{align*}
 \mathcal{L}(\bm{r}_d, \bm{b}_d) = \sum_{i=1}^{m_d} \mathcal{L}(\bm{b}_d[i] - \bm{r}_d[i]).
 \end{align*}
 with $\bm{b}_d \in \mathbb{R}^{m_d}$.
 
Since $\mathcal{L}$ and $\| \ \|_2^2$ are both additive, we can  divide the optimization problem (\ref{rupdate}) into $m_d$  independent univariate problems. Then,  $\bm r_d^{k+1}$ is obtained component-wise and can be expressed as
\begin{equation}\small
\label{Crupdate}
\begin{aligned}
\bm r_d^{k+1}[i] \ =&\mathop {\arg \min }\limits_{\bm r_d[i]} \{ \mathcal{L}(\bm{r}_d[i], \bm{b}_d[i])  + \frac{\mu}{2}  \| \bm A_d[i,] \bm{x}^{k+1}  \\
 &  - {\bm{r}}_d[i]  - {\bm{u}}^k_d[i]/\mu  \|_2^2, \}, i=1.\dots,n_m.
\end{aligned}
\end{equation}
 Regardless of whether $\mathcal{L}$  is used for classification or regression problems, solving the above subproblem is straightforward. We will delve into this in Section \ref{sec5}, elucidating its application in some specific cases.

In summary, we present the detailed process of the proposed parallel framework in Algorithm \ref{alg1} and visualize it in Figure \ref{fig2}. Compared to the consensus parallel algorithm in Figure \ref{fig1}, which requires updating $(2D+1)n$ variables, our proposed parallel algorithm framework only requires updating $m+n$ variables, which is independent of the number of slave nodes $D$. For communication between master and slave nodes, our communication cost is also lower. Specifically, our algorithm only needs to pass on an $m_d$-dimensional vector from the master node to the slave nodes, while the consensus parallel ADMM needs to pass on an $n$-dimensional vector. This has a significant advantage when the number of features $n$ is particularly large.

\begin{algorithm}\small
\caption{\small{Partition-insensitive Parallel Framework for Distributed Model Fitting}}
\label{alg1}
\begin{algorithmic}
\STATE {\textbf{Pre-computation}:  $\eta_d =\text{eigen}(\mu \bm A_d^\top \bm A_d )$, $d= 1,2,\dots,D$.} 
\STATE {\textbf{Input:} $\bullet$ Master node: $\mathcal{R}(\bm{x})$, $\mu$ and $\bm{x}^{0}$.\\
\qquad \  \ \ \  $\bullet$ The $d$-th slave node:  $\{\boldsymbol{A}_d,\boldsymbol{b}_d\}$; $\mu$, $\mathcal{L}$ and $\bm u_d^0$.}
\STATE {\textbf{Output:} the total number of iterations $K$,  $\boldsymbol{x}^K$, $\boldsymbol{r}^K$, $\boldsymbol{u}^K$. }
\STATE {\textbf{while} not converged \textbf{do}}
\STATE {\textbf{Master node}: 1. Receive $\eta_d$ (only once) and $\bm \xi_d^k$ posted by $M$ slave nodes, \\
\qquad  \qquad \quad \ \ \ \ 2. Update $\bm x^{k+1}$ using (\ref{xxx}), \\
\qquad  \qquad \quad \ \ \ \ 3. Send $\bm x^{k+1}$ to each slave node. 
}
\STATE {\textbf{Slave nodes}: \ \  for $d =1 ,2, \dots, D$ (in parallel) \\
\qquad  \qquad \quad \ \ \ \ 1. Receive $\bm x^{k+1}$  posted by master node, \\
\qquad  \qquad \quad \ \ \ \ 2. Update $\bm r_d^{k+1}$  using (\ref{Crupdate}), \\
\qquad  \qquad \quad \ \ \ \ 3. Update $\boldsymbol u_d ^{k + 1}  =   \bm u_d ^k - \mu ( \bm A_d \bm x^{k+1}  - \bm r_d^{k+1} )$, \\
\qquad  \qquad \quad \ \ \ \ 4. Compute $\bm \xi_d^{k+1} =  \bm A_d^\top ({\bm{r}_d^{k+1}}  + {\bm{u}_d^{k+1}}/\mu)$, \\
\qquad  \qquad \quad \ \ \ \ 5. Send $\bm \xi^{k+1}_d$  to master node.
}
\STATE {\textbf{end while}}
\STATE {\textbf{return} solution}.
\end{algorithmic}
\end{algorithm}

\section{Theoretical Properties of Algorithm \ref{alg1}}\label{sec4}
In this section, we theoretically prove that Algorithm \ref{alg1} has two properties: one is partition insensitivity, and the other is that the iterative solution of the algorithm has a linear convergence rate.

\subsection{Partition insensitivity}\label{sec4.1}
Let $D_1$ and $D_2$ be two different numbers of slave nodes, i.e. $D_1 \ne D_2$.
For ease of distinction,  let us denote  $\{ \hat{\bm x}^k, \hat{\bm r}^k, \hat{\bm u}^k\}$ as the $k$-th iteration results of Algorithm  \ref{alg1} with $D_1$ slave nodes, and  $\{ \tilde{\bm x}^k, \tilde{\bm r}^k , \tilde{\bm u}^k \}$ as the $k$-th iteration results of Algorithm  \ref{alg1} with $D_2$ slave nodes. Here, we assume that  the above two situations use the same $\eta$ and $\{ \hat{\bm x}^k, \hat{\bm r}^k, \hat{\bm u}^k\} = \{ \tilde{\bm x}^k, \tilde{\bm r}^k , \tilde{\bm u}^k \} $.

Next, we will discuss the solutions for $k+1$ iteration in two situations. According to (\ref{xixi}),  we have 
\begin{equation}
\begin{aligned}
&\sum_{d=1}^{D_1} \bm A_d^\top (\bm A_d \hat{\bm x}^k  - {\hat{\bm{r}}_d^k}  - {\hat{\bm{u}}_d^k}/\mu)\\
= &\sum_{d=1}^{D_2} \bm A_d^\top (\bm A_d \tilde{\bm x}^k  - \tilde{\bm r}^k  - \tilde{\bm u}^k/\mu). 
\end{aligned}
\end{equation}
Together with $\{ \hat{\bm x}^k, \hat{\bm r}^k, \hat{\bm u}^k\} = \{ \tilde{\bm x}^k, \tilde{\bm r}^k , \tilde{\bm u}^k \}$, we can conclude $\hat{\bm x}^{k+1} = \tilde{\bm x}^{k+1}$ from the update of $\bm x^{k+1}$-update in (\ref{xxx}). From (\ref{Crupdate}), we can easily derive $\hat{\bm r}^{k+1} = \tilde{\bm r}^{k+1} $  from $\hat{\bm x}^{k+1} = \tilde{\bm x}^{k+1}$ and $\hat{\bm u}^k = \tilde{\bm u}^k$. Looking back at the update methods of dual variables $\boldsymbol u_d ^{k + 1}  =   \bm u_d ^k - \mu ( \bm A_d \bm x^{k+1}  - \bm r_d^{k+1} )$, we can conclude that  $\hat{\bm u}^{k+1} = \tilde{\bm u}^{k+1} $. Therefore, the conclusion that $\{ \hat{\bm x}^{k+1}, \hat{\bm r}^{k+1}, \hat{\bm u}^{k+1}\} = \{ \tilde{\bm x}^{k+1}, \tilde{\bm r}^{k+1} , \tilde{\bm u}^{k+1} \} $ is confirmed.

From the above discussion,  we can draw the following  surprising conclusion.
\begin{thm}\label{TH1}
If we use the same initial  iteration variables $$\{ \hat{\bm x}^0, \hat{\bm r}^0, \hat{\bm u}^0 \} = \{ \tilde{\bm x}^0, \tilde{\bm r}^0 , \tilde{\bm u}^0 \}$$ and the linearized parameter $\eta$,  the iterative  solutions  obtained  by Algorithm \ref{alg1} with different slave nodes are actually the same, i.e.,
\begin{align}
\left\{ \hat{\bm x}^k, \hat{\bm r}^k, \hat{\bm u}^k\right\} = \left\{ \tilde{\bm x}^k, \tilde{\bm r}^k , \tilde{\bm u}^k \right\}, \ \text{for} \ \text{all} \  {k}.
\end{align}
\end{thm}

This theorem indicates if Algorithm \ref{alg1} begins with the same initial values and $\eta$, its iterative solution remains unaffected regardless of how samples are partitioned. This phenomenon is referred to as sample partitioning insensitivity. In other words, maintaining a consistent $\eta$ ensures that variations in $D$ and the number of samples processed by each node do not impact the iterative solution of the parallel algorithm.

However, as $D$ increases, $\eta=\sum_{d=1}^D \eta_d$ will also increase. Several studies have investigated the influence of the parameter $\eta$ in the linearized ADMM algorithm. They have concluded that a large $\eta$ value leads to slower convergence of the algorithm, with potentially insignificant alterations in the iterative convergence solution. Interested readers can delve into the research conducted by \cite{He2020Optimally} and its associated references. Subsequently, we will develop the convergence analysis of the algorithm.

\subsection{Linear convergence rate}\label{sec4.2}
According to Theorem \ref{TH1}, as long as $\eta$ is the same, the solution of Algorithm \ref{alg1} is the same regardless of how many slave nodes there are. Therefore, we only need to discuss the convergence of the iterative solution of Algorithm \ref{alg1} in the case of $D=1$. At this point, the scheme of algorithm is defined as
\begin{footnotesize}
\begin{equation*}
\label{primalupdate}
\left\{ \begin{aligned}
\bm x^{k+1}  = \ & \mathop {\arg \min }\limits_{\bm x} \{ \mathcal{R}(\bm{x}) + \frac{\mu}{2} \| \bm A \bm{x} - {\bm{r}^k}  - {\bm{u}^k}/\mu  \|_2^2 \\
&\qquad \qquad +  \frac{1}{2} \left\| \bm x - \bm x^k  \right\|_{\bm S}^2 \}, \\%
\bm r^{k+1}  = \ & \mathop {\arg \min }\limits_{\bm r} \{ \mathcal{L}(\bm{r}, \bm{b}) + \frac{\mu}{2}  \left\| \bm A \bm{x}^{k+1} - {\bm{r}}  - {\bm{u}}^k/\mu  \right\|_2^2 \},\\%
\boldsymbol u ^{k + 1} = \ &  \  \bm u ^k - \mu (\bm A \bm{x}^{k+1} - {\bm{r}^{k+1}} ).
\end{aligned} \right.
\end{equation*}
\end{footnotesize}
If  $\mathcal{R}$  is convex with respect to $\bm x$ and $\mathcal{L}$ is convex with respect to $\bm r$, the convergence properties of the above iterative schemes have been widely studied, see\cite{Liang2024Linearized}, \cite{He2012On}  and \cite{He2015On}. The convergence of Algorithm \ref{alg1} is described as follows.
\begin{thm}\label{TH2}
For the $\eta > \text{eigen}(\mu \bm A^\top \bm A )$,
the sequence $\boldsymbol{w}^{k}=\{\bm x^k, \boldsymbol{r}^k, \boldsymbol{u}^k\}$ is generated by Algorithms \ref{alg1} with an arbitrary initial feasible solution $\boldsymbol{w}^{0}$. Then the sequence $\boldsymbol{w}^{k}$ converges to $\boldsymbol{w}^{*}=\{\bm x^*, \boldsymbol{r}^*, \boldsymbol{u}^*\}$, where  $\boldsymbol{w}^{*}$ is an arbitrary optimal solution point of the (\ref{sec3eq1}). The $O(1/k)$ convergence rate in a non-ergodic sense can be obtained, i.e.,
\begin{equation}
\| \boldsymbol{w}^{k}-\boldsymbol{w}^{k+1} \|_{\boldsymbol{H}}^{2} \le \small{\frac{1}{k+1}}\| \boldsymbol{w}^{0}-\boldsymbol{w}^{*} \|_{\boldsymbol{H}}^{2},
\end{equation}
where $\boldsymbol{H}=\left[\begin{array}{*{20}{c}}
\eta \bm I_{n}-{\bm A}^\top  {\bm A} & \bm 0& \bm 0  \\
 \bm 0 &  \mu \bm I_m &  \bm0 \\ 
\bm 0 & \bm 0 &\frac{1}{\mu}\bm I_m
\end{array} \right] $ is a symmetric and positive matrix ($\bm I_m$ an m-dimensional identity matrix). 
\end{thm}

\section{Application of Parallel Frameworks}\label{sec5}

As \cite{Boyd2010Distributed} showed, one of the most important applications of model fitting is are classification  and regression problems. 
 Next, we will use our designed parallel algorithms to solve some popular classification and regression problems. When considering the existence of intercept terms, we can set all the elements in the first column of  the feature matrix $\bm A$ to 1.

\subsection{Classification problems}\label{sec5.1}
Here, we mainly focus on two classification models, one is regularized logistic regression model \cite{Boyd2010Distributed,Yang2015A} and the other is regularized support vector machines \cite{Boyd2010Distributed,Liang2024Linearized}.

As introduced in Section \ref{sec3.2}, the $\mathcal{L}$-function for classification problems is
$$\mathcal{L}(\bm{r}, \bm{b}) = \sum_{d=1}^{D} \mathcal{L}(\bm{r}_d, \bm{b}_d) = \sum_{d=1}^{D} \sum_{i=1}^{m_d}\mathcal{L}(\bm{r}_d[i] \bm{b}_d[i]).$$
Then, classification models can be formulated as
\begin{equation}\label{clamodel}
\begin{aligned}
&\min_{\bm{x},\{\bm{r}_d\}} {{\kern 1pt} {\kern 1pt} {\kern 1pt} {\kern 1pt} \sum_{d=1}^{D} \sum_{i=1}^{m_d}\mathcal{L}(\bm{r}_d[i] \bm{b}_d[i]) +  \mathcal{R}(\bm{x})}\\
&\text{s.t.}{\kern 1pt} {\kern 1pt} {\kern 1pt} {\kern 1pt} {\kern 1pt} {\kern 1pt} \bm{A}_d\bm x = \bm{r}_d, d=1,2,\dots,D. 
\end{aligned}
\end{equation}

$\bullet$ \textbf{Logistic regression}. The logistic loss is
$$\mathcal{L}(\bm{r}_d[i] \bm{b}_d[i])  = \log(1 + \exp^{-\bm{r}_d[i] \bm{b}_d[i]}). $$
Then,  the $\bm x^{k+1}$-update in Algorithm \ref{alg1} is
$$\bm x^{k+1} = \text{prox}_{\gamma, \mathcal{R}(\bm{x})}(\bm x^k   - \frac{\mu}{\eta} \sum_{d=1}^D \bm \xi_d^k),$$
and the $\bm r_d^{k+1}[i]$-update in Algorithm \ref{alg1} is 
\begin{equation}\label{Logirupdate}
\begin{split}
\mathop {\arg \min }\limits_{\bm r_d[i]}& \left\{ \log(1 + \exp^{-(\bm{r}_d[i] \bm{b}_d[i])}) + \frac{\mu}{2}  \| \bm A_d[i,] \bm{x}^{k+1} \right. \\ &\left.  - {\bm{r}}_d[i]  - {\bm{u}}^k_d[i]/\mu  \|_2^2 \right\}, i=1.\dots,n_m.
	\end{split}
\end{equation}
The above  formula has no closed-form solution and can be effectively solved using Newton's method. The first and second derivatives of $\mathcal{L}(\bm{r}_d[i] \bm{b}_d[i])$ are defined as follows, respectively,
\begin{equation}\label{der}
\begin{split}
  \mathcal{L}^{'}(\bm{r}_d[i] \bm{b}_d[i]) = \frac{-\bm{b}_d[i]}{\exp^{\bm{r}_d[i] \bm{b}_d[i]}+1},  \\
  \mathcal{L}^{''}(\bm{r}_d[i] \bm{b}_d[i]) =  \frac{\exp^{\bm{r}_d[i] \bm{b}_d[i]}}{(\exp^{\bm{r}_d[i] \bm{b}_d[i]}+1)^2}. 
  \end{split}
\end{equation}
Therefore, $\bm r_d^{k+1}[i]$ can be obtained by using the second-order  Newton iterations in Algorithm \ref{alg2} until convergence. 
\begin{algorithm}\small
\caption{\small{Newton's method for $\bm r_d^{k+1}[i]$-update  }}
\label{alg2}
\begin{algorithmic}
\STATE {1. Initialize $\bm r_d^{k+\frac{1}{2},0}[i]$ with  $\bm r_d^{k}[i]$, where $\bm r_d^{k}[i]$ is obtained from the last ADMM iteration.}
\STATE {2. For $l=0,1,2,\dots,L$, continue iterating the Newton iteration until convergence is achieved. }
\STATE {\quad  2.1. Compute $\mathcal{L}^{'}(\bm{r}_d[i]^{k+\frac{1}{2},l} \bm{b}_d[i])$ and $\mathcal{L}^{''}(\bm{r}_d[i]^{k+\frac{1}{2},l} \bm{b}_d[i])$ by (\ref{der}),}
\STATE {\quad 2.2. Compute $\bm r_d^{k+\frac{1}{2},l+1}[i] = \Big[\mu(\bm A_d[i,] \bm{x}^{k+1} - {\bm{u}}^k_d[i]/\mu) + \bm{r}_d[i]^{k+\frac{1}{2},l} \mathcal{L}^{''}(\bm{r}_d[i]^{k+\frac{1}{2},l}\bm{b}_d[i]) -\mathcal{L}^{'}(\bm{r}_d[i]^{k+\frac{1}{2},l} \bm{b}_d[i]))\Big]/{(\mathcal{L}^{''}(\bm{r}_d[i]^{k+\frac{1}{2},l} \bm{b}_d[i]) + \mu)} $, }
\STATE {\quad  2.3. If the solution converges, then define $\bm r_d^{k+1}[i] = \bm r_d^{k+\frac{1}{2},l+1}[i]$.}
\end{algorithmic}
\end{algorithm}

Our method for solving the subproblem $\bm r$ is somewhat similar to the second-order Newton iteration algorithm used in \cite{Lin2007Trust}, but please note that $\bm Ax$ has been replaced by $\bm r$. At this point, the solution problem is a regular quadratic optimization problem that can converge quickly. This will significantly improve the efficiency of ADMM algorithm in solving regularized logistic regression, as demonstrated by our simulation experiments in Section \ref{sec6.1}.

$\bullet$ \textbf{Support vector machines}. \cite{Liang2024Linearized} indicated that many SVM loss functions, such as hinge loss, squared hinge loss,  huberized  hinge loss,  pinball loss,  huberized pinball loss and  asymmetric least squares loss, can be expressed as 
$$\mathcal{L}(\bm{r}_d[i] \bm{b}_d[i]) = \mathcal{L}(1 - \bm{r}_d[i] \bm{b}_d[i]),$$
where $\mathcal{L}$ represents the corresponding loss function mentioned above. For example, for hinge loss,  $\mathcal{L}(1 - \bm{r}_d[i] \bm{b}_d[i])=(1 - \bm{r}_d[i] \bm{b}_d[i])_+$; for squared hinge loss, $\mathcal{L}(1 - \bm{r}_d[i] \bm{b}_d[i])=\frac{1}{2}(1 - \bm{r}_d[i] \bm{b}_d[i])_+^2$.

Similarly,  the $\bm x^{k+1}$-update in Algorithm \ref{alg1} is
$$\bm x^{k+1} = \text{prox}_{\gamma, \mathcal{R}(\bm{x})}(\bm x^k - \frac{\mu}{\eta}\sum_{d=1}^D \bm \xi_d^k ),$$
and the $\bm r_d^{k+1}[i]$-update in Algorithm \ref{alg1} is 
\begin{equation}\label{svmupdate}
\begin{split}
\mathop {\arg \min }\limits_{\bm r_d[i]}& \left\{ \mathcal{L}(1 - \bm{r}_d[i] \bm{b}_d[i]) + \frac{\mu}{2}  \| \bm A_d[i,] \bm{x}^{k+1} \right. \\ &\left.  - {\bm{r}}_d[i]  - {\bm{u}}^k_d[i]/\mu  \|_2^2 \right\}, i=1.\dots,n_m.
	\end{split}
\end{equation}
Through simple variable substitution and algebraic operations, we can  demonstrate that the above formula can be transformed into a linear function of the $\mathcal{L}$ proximal operator in \cite{Liang2024Linearized}, that is,
\begin{equation}
\begin{aligned}
\bm r_d^{k+1}[i] = & \bm{b}_d[i] - \bm{b}_d[i] \cdot \operatorname{prox}_{\mu, \mathcal{L}}(1- \bm{b}_d[i]\bm A_d[i,] \bm{x}^{k+1}\\
& + \bm{b}_d[i]\bm{u}^k_d[i]/\mu ) . 
\end{aligned}
\end{equation}

\subsection{Regression Problems}\label{sec5.2}
In this section, we apply Algorithm \ref{alg1} to solving regression problems.
As introduced in Section \ref{sec3.2},  $\mathcal{L}$-function for regression problems is
$$\mathcal{L}(\bm{r}, \bm{b}) = \sum_{d=1}^{D} \mathcal{L}(\bm{r}_d, \bm{b}_d) = \sum_{d=1}^{D} \sum_{i=1}^{m_d}\mathcal{L}( \bm{b}_d[i] -\bm{r}_d[i]).$$
Then, regression models can be formulated as
\begin{align}\label{Rlamodel}
\min_{\bm{x},\{\bm{r}_d\}} {{\kern 1pt} {\kern 1pt} {\kern 1pt} {\kern 1pt} \sum_{d=1}^{D} \sum_{i=1}^{m_d}\mathcal{L}(\bm{b}_d[i] - \bm{r}_d[i] ) +  \mathcal{R}(\bm{x})}\\
\text{s.t.} {\kern 1pt} {\kern 1pt} {\kern 1pt} {\kern 1pt} \bm{A}_d\bm x = \bm{r}_d, d=1,2,\dots,D. \notag
\end{align}
Many loss regression losses can be incorporated into the above optimization formulas, that is, $\mathcal{L}$ can be  least squares loss, quantile loss, smooth quantile loss (two types), Huber loss, and SVR (support vector regression) loss and so on. The closed-form solutions for the proximal operators mentioned above (excluding SVR) have been derived. Further details can be referenced in \cite{Wu2023Partition} and \cite{Wu2024Multi}. Moreover, a proof of the proximal operator for the subproblem of SVR is provided in appendix \ref{C}.

Similarly, the $\bm x^{k+1}$-update in Algorithm \ref{alg1} is
$$\bm x^{k+1} = \text{prox}_{\gamma, \mathcal{R}(\bm{x})}(\bm x^k - \frac{\mu}{\eta}\sum_{d=1}^D \bm \xi_d^k ),$$
and the $\bm r_d^{k+1}[i]$-update in Algorithm \ref{alg1} is 
\begin{equation}\label{regupdate}
\begin{split}
\mathop {\arg \min }\limits_{\bm r_d[i]}& \left\{ \mathcal{L}(\bm{b}_d[i] - \bm{r}_d[i] ) + \frac{\mu}{2}  \| \bm A_d[i,] \bm{x}^{k+1} \right. \\ &\left.  - {\bm{r}}_d[i]  - {\bm{u}}^k_d[i]/\mu  \|_2^2 \right\}, i=1.\dots,n_m.
	\end{split}
\end{equation}
Through simple variable substitution and algebraic operations, we can  demonstrate that the above formula can be transformed into a linear function of the $\mathcal{L}$, that is,
\begin{small}
$$\bm r_d^{k+1}[i] = \bm{b}_d[i] - \operatorname{prox}_{\mu, \mathcal{L}}(\bm{b}_d[i] - \bm A_d[i,] \bm{x}^{k+1} + \bm{u}^k_d[i]/\mu ). $$
\end{small}

\section{Numerical experiment}\label{sec6}
The resources we uploaded include implementation codes for various models in Section \ref{sec5}, as well as some simulation experiments on some synthetic data.  These models include $\ell_1, \ell_2, \ell_{2,1}$ regularized logistic regression; $\ell_1, \ell_2, \ell_{2,1}$ regularized SVM classification models with six types of losses in Section \ref{sec5.1};  and $\ell_1, \ell_2, \ell_{2,1}$ regularized regression models  with six types of losses in Section \ref{sec5.2}. In this section, we only present the experimental results of real-world data. All experiments were performed using R on a computer equipped with an AMD Ryzen 9 7950X 16-Core Processor running at 4.50 GHz and with 64 GB RAM. For the convenience of readers to verify the effectiveness of algorithms on a single machine, we provide pseudo parallel code for one machine.

To choose the optimal values for the regularization parameters $\lambda$, we follow the approach proposed by \cite{Lee2014Model}. We minimize the HBIC criterion, defined as
\begin{equation}
\begin{aligned}
\text{HBIC}(\lambda) = &\log\left(\sum_{i=1}^m \mathcal{L}_i(\bm{a}_i^\top \hat{\bm{x}}_{\lambda}, b_i)\right)\\
 &+ |S_{\lambda}| \frac{\log(\log m)}{m} C_m.
\end{aligned}
\end{equation}
Here, $\mathcal{L}$ represents a specific loss function, and $\hat{\bm{x}}_{\lambda}$ corresponds to the estimator obtained. $|S_{\lambda}|$ denotes the number of nonzero coordinates in $\hat{\bm{x}}_{\lambda}$, and the value $C_n = 6 \log (n)$ is recommended by \cite{Peng2015An}.
By minimizing the HBIC criterion, we can effectively select the optimal $\lambda$  values for our  estimators. These choices allow us to balance the trade-off between model complexity and goodness of fit.

For all ADMM algorithms, we set the maximum iteration number to 500, with the stopping criterion defined as follows:
\begin{align*}
\frac{\|\bm w^k - \bm w^{k-1} \|_2}{\max (1, \|\bm w^k \|_2)} \le 10^{-2}.
\end{align*}
This stopping criterion is consistent with the theoretical result of Theorem \ref{TH2}, and ensures that the difference between consecutive iterations of the estimated coefficients does not exceed a specified threshold. In the following numerical simulation, we conduct regression experiments using synthesized data and classification experiments using a real dataset.
\subsection{Synthetic data}
We utilized simulated models in the simulation studies of \cite{Fan2021Penalized}. Specifically, data were generated from the heteroscedastic regression model $y = x_6 + x_{12} + x_{15} + x_{20} + 0.7 x_1 \epsilon$, where $\epsilon \sim N(0,1)$. The covariates $(x_1,x_2,\dots,x_p)$ are generated in two steps.

\begin{itemize}
  \item Firstly, we generated $\bm{\tilde{x}} = (\tilde{x}_1, \tilde{x}_2, \dots, \tilde{x}_p)^\top$ from a $p$-dimensional multivariate normal distribution $N(\bm{0}, \bm{\Sigma})$, where $\Sigma_{ij} = 0.5^{|i-j|}$ for $1 \le i, j \le p$.
  
  \item Secondly, we set $x_1 = \Phi(\tilde{x}_1)$ and $x_j = \tilde{x}_j$ for $j=2,\dots,p$.
\end{itemize}
We set $n = 10000$ and $p = 20000$, and use six regression losses (least squares, quantile, smooth quantile(two types), Huber, and SVR) with $\ell_1$ regularization terms to fit this data. 

We record the number of iterations (NI), calculation time (CT), false negative (FN,  non-zero coefficient estimation is zero),  false positive (FP, zero coefficient estimation is non-zero),  mean absolute error (MAE), and mean square error (MSE) to measure the performance of each algorithm.

\begin{table*}[!ht]\footnotesize
    \centering
    \caption{Comparison between PIPADMM and consensus parallel ADMM algorithms.}
    \renewcommand{\arraystretch}{1.2}
    \resizebox{\textwidth}{!}{
    \begin{tabular}{cccccccccccccc}
    \hline
    & & \multicolumn{6}{l}{PIPADMM} & \multicolumn{6}{l}{CPADMM}\\
    \cmidrule(lr){3-8}\cmidrule(lr){9-14}
    Model & D & NI & CT & FN & FP & MAE & MSE & NI & CT & FN & FP & MAE & MSE \\
    \hline
    \multirowcell{3}{Lasso}& 1 & 15.2 & 63.96 & 0 & 0 & 0.2824 & 0.4088 & 89.3 & 1835.5 & 0 & 2.9 & 0.4412 & 0.7985 \\
                           & 5 & 14.8 & 17.33 & 0 & 0 & 0.2956 & 0.4172 & 98.6 & 864.7 & 0 & 4.6 & 0.4236 & 0.7514 \\
                           & 10 & 15.7 & 9.625 & 0 & 0 & 0.3123 & 0.4319 & 113.8 & 496.5 & 0 & 6.2 & 0.4099 & 0.7436 \\
    \hline
    \multirowcell{3}{$\ell_1$-quantile}& 1 & 30.1 & 105.6 & 0 & 0 & 0.2812 & 0.4079 & 156.3 & 2269.7 & 0 & 3.2 & 0.4329 & 0.7823 \\
                                       & 5 & 31.7 & 59.73 & 0 & 0 & 0.2810 & 0.4077 & 169.8 & 1365.5 & 0 & 4.8 & 0.4276 & 0.7661 \\
                                       & 10 & 32.3 & 20.21 & 0 & 0 & 0.2809 & 0.4074 & 183.3 & 522.6 & 0 & 6.1 & 0.4195 & 0.7530 \\
    \hline
    \multirowcell{3}{$\ell_1$-Huber}& 1 & 36.0 & 123.9 & 0 & 0 & 0.2808 & 0.4055 & 176.3 & 2558.0 & 0 & 4.6 & 0.4562 & 0.7912 \\
                                    & 5 & 35.9 & 68.84 & 0 & 0 & 0.2911 & 0.4062 & 189.7 & 1497.6 & 0 & 4.9 & 0.4427 & 0.7836 \\
                                    & 10 & 36.5 & 29.77 & 0 & 0 & 0.2965 & 0.4069 & 211.9 & 667.3 & 0 & 5.8 & 0.4288 & 0.7796 \\
    \hline
    \end{tabular}}
    \label{tab3}
\end{table*}
The consensus ADMM parallel algorithms for Lasso and $\ell_1$-Huber regression were proposed by \cite{Boyd2010Distributed}, while the consensus ADMM parallel algorithm for $\ell_1$-quantile regression was initially presented in \cite{Yu2017A} and later enhanced by \cite{Fan2021Penalized}. In Table \ref{tab3}, we offer a comparison between our partition-insensitive ADMM parallel algorithm (PIPADMM) and the consensus parallel ADMM (CPADMM) algorithm. The results from the numerical experiments suggest that PIPADMM outperforms CPADMM significantly in terms of algorithm computation speed, accuracy, and variable selection.

Recently, \cite{Wu2023Partition} introduced a partition-insensitive parallel ADMM algorithm. To distinguish and enable easy comparison, in Table \ref{tab4}, we label our algorithm as PIPADMM1 and their algorithm as PIPADMM2. Experimental results indicate that the performance of the two parallel algorithms across various regression models is generally comparable. Furthermore, it is important to note that $\ell_1$-SVR can solely be addressed by the parallel algorithm proposed in this paper; thus, the experimental results presented are solely for our algorithm.
\begin{table*}[htb]\footnotesize
    \centering
    \caption{Comparison of two partition insensitive parallel algorithms.}
    \renewcommand{\arraystretch}{1.3}
    \resizebox{\textwidth}{!}{
    \begin{tabular}{cccccccccccccc}
    \hline
    & & \multicolumn{6}{l}{PIPADMM1} & \multicolumn{6}{l}{PIPADMM2}\\
    \cmidrule(lr){3-8}\cmidrule(lr){9-14}
    Model & D & NI & CT & FN & FP & MAE & MSE & NI & CT & FN & FP & MAE & MSE \\
    \hline
    \multirowcell{3}{$\ell_1$-SQ1}& 1 & 25.1 & 91.83 & 0 & 0 & 0.2852 & 0.4086 & 25.5 & 92.36 & 0 & 0 & 0.2863 & 0.3997\\
                                  & 5 & 25.5 & 63.76 & 0 & 0 & 0.2907 & 0.4104 & 24.9 & 62.12 & 0 & 0 & 0.2894 & 0.4038 \\
                                  & 10 & 26.3 & 41.25 & 0 & 0 & 0.2966 & 0.4172 & 25.8 & 40.97 & 0 & 0 & 0.2961 & 0.4071 \\
    \hline
    \multirowcell{3}{$\ell_1$-SQ2}& 1 & 27.2 & 101.8 & 0 & 0 & 0.2830 & 0.4083  & 27.6 & 103.5 & 0 & 0 & 0.2826 & 0.4103 \\
                                  & 5 & 27.3 & 37.19 & 0 & 0 & 0.2834 & 0.4089  & 27.2 & 36.27 & 0 & 0 & 0.2840 & 0.4077 \\
                                  & 10 & 27.1 & 21.33 & 0 & 0 & 0.2858 & 0.4056  & 27.4 & 21.58 & 0 & 0 & 0.2855 & 0.4094 \\
    \hline
    \multirowcell{3}{$\ell_1$-SVR}& 1 & 63.3 & 94.25 & 0 & 0 & 0.999 & 1.2532  & - & - & - & - & - & - \\
                                  & 5 & 68.2 & 67.52 & 0 & 0 & 0.955 & 1.1971  & - & - & - & - & - & - \\ 
                                  & 10 & 73.5 & 49.89 & 0 & 0 & 0.971 & 1.2185  & - & - & - & - & - & - \\
    \hline
    \end{tabular}}
    \label{tab4}
\end{table*}

\subsection{Real data}
\subsubsection{Data introduction and algorithm implementation details}\label{sec6.0}
 One of the key distinctions between this paper and the work by \cite{Wu2023Partition} is our ability to employ regularized logistic regressions and regularized SVMs for data classification. Therefore, the simulation experiments in this section primarily center on classification data.
We use the rcv1.binary dataset, which has 47236 features, 20242 training samples, and 677399 testing samples. It can be publicly obtained through the website \url{https://www.csie.ntu.edu.tw/~cjlin/libsvmtools/datasets/binary.html#rcv1.binar}. 

In the following experiment, we trained the model using training samples, where the dimension of data $\bm A$ is $m=20242$ and $n=47236$, indicating that $n>m$ is a high-dimensional dataset. Considering the large number of data features we use, many of which do not contribute to the classification of data categories, we use $\ell_1$ regularization terms. For this dataset, we recommend using $\mu=0.1$, which can also be selected from the candidate set through cross validation. For initial values $\bm x^0$ and $\bm u_d^0$, we set all their elements to 0.1, and in fact, any other value is also acceptable.

Next, we will describe some evaluation metrics for the algorithm. For the test set data, we will randomly sample 10000 from 677399 testing samples, which means $n_{test}=10000$. “Time” represents the running time of the algorithm, “Iteration” represents the number of iterations of the algorithm, “Sparsity”  represents the proportion of 0 coefficients (estimated as 0 coefficients divided by the total number of coefficients) , “Train” represents the accuracy of estimating coefficient classification on the training set, and “Test” represents the accuracy of estimating coefficient classification on the testing set. 

\subsubsection{Regularized logistic regression}\label{sec6.1}
\begin{figure*}[h]\centering
	\includegraphics[width=14cm]{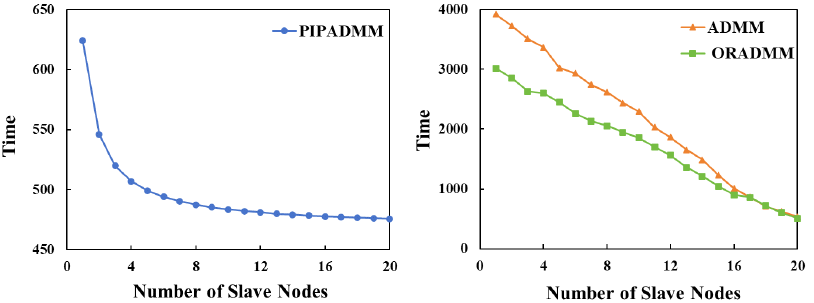}
	\caption{Illustration of the variation in training time for $\ell_1$ logistic regression with three parallel algorithms as the number of slave nodes changes.}
    \label{fig3}
\end{figure*}

Here, we compare the  parallel ADMM algorithms in  \cite{Boyd2010Distributed} (ADMM) and \cite{Qiu2023PSRA} (PSRA-HGADMM) with our algorithm (PIPADMM) for calculating $\ell_1$ logistic regression in \cite{Park2007L1}. Due to the limited running memory of the computer, we can only achieve a maximum of 20 parallel slave nodes. In a parallel environment, our data is divided in order and in the same proportion, resulting in a higher or lower number of remaining copies.
The performance of three parallel algorithms on different slave nodes is shown in Figure \ref{fig3} and Table \ref{tab1}. 

\begin{table*}[!ht]\footnotesize
    \centering
    \renewcommand{\arraystretch}{1.2}
    \caption{Comparative analysis of Sparsity,  training,  and average testing accuracies (\%) for $\ell_1$ logistic regression using PIPADMM, ADMM, and PSRA-HGADMM}
    \begin{tabular}{llllllllll}
    \hline
          & \multicolumn{3}{c}{PIPADMM} & \multicolumn{3}{c}{ADMM} & \multicolumn{3}{c}{PSRA-HGADMM}\\ 
        \cmidrule(lr){2-4}\cmidrule(lr){5-7}\cmidrule(lr){8-10}
        D & Sparsity & Train & Test & Sparsity & Train & Test & Sparsity & Train & Test \\ \hline
        2 & 86.47 & 90.24 & 87.36 & 80.64 & 88.25 & 85.36 & 85.64 & 90.36 & 88.72 \\ 
        4 & 86.40  & 90.19 & 87.21 & 79.25 & 86.67 & 84.12 & 84.25 & 89.02 & 86.88 \\ 
        6 & 86.35 & 89.77 & 87.13 & 78.56 & 85.97 & 83.72 & 83.69 & 88.45 & 85.97 \\ 
        8 & 86.32 & 89.13 & 86.99 & 77.36 & 85.14 & 82.91 & 82.28 & 87.09 & 84.82 \\ 
        10 & 85.30  & 88.72 & 86.27  & 76.36 & 84.55 & 81.67 & 81.58 & 85.59 & 84.06 \\ 
        12 & 84.25 & 88.09 & 86.05 & 75.78 & 84.01 & 81.08 & 80.69 & 84.13 & 83.21 \\ 
        14 & 84.21 & 87.71 & 85.81 & 74.98 & 83.36 & 80.76 & 79.85 & 83.08 & 82.44 \\ 
        16 & 84.15 & 87.14 & 85.36 & 73.68 & 82.99 & 80.29 & 78.58 & 81.97 & 81.26 \\ 
        18 & 84.10  & 86.69 & 85.09 & 73.01 & 82.11 & 79.77 & 77.54 & 81.26 & 80.28 \\ 
        20 & 83.85 & 86.92 & 84.81 & 71.68 & 81.56 & 79.15 & 76.36 & 80.75 & 79.69 \\ \hline
    \end{tabular}
    \label{tab1}
\end{table*}

Figure \ref{fig3} 
shows the time required for training the $\ell_1$ logistic regression of three parallel algorithms at different slave nodes. Obviously, our algorithm requires much less time than the other two parallel algorithms. However, as the number of slave nodes increases, our algorithm training time does not decrease as much as the other two consensus ADMM algorithms. The main reason for this phenomenon is that in ADMM and PSRA-HGADMM algorithms, besides the time-consuming update of large-scale $\bm x$, the Newton iteration of the $\bm z_d$-sub problem in (\ref{cz})  is also time-consuming due to the presence of the $\bm A_d$ matrix, resulting in a large number of large-scale matrices and vector multiplication. Fortunately, as the number of nodes increases, the row of $\bm A_d$  rapidly decreases, so the training time of these two algorithms is greatly affected by the number of nodes. 

However, our algorithm $\bm r_d$ is just an ordinary quadratic optimization problem, and the second-order Newton iteration method converges quickly without much computational burden. Increasing the number of nodes only reduces a portion of the optimization time for $\bm r_d$. 
 In fact, the main computational burden of PIPADMM in solving $\ell_1$ logistic regression here lies in the update of the large dimension $\bm x$. The increase in the number of slave nodes does not alleviate its computational burden. Therefore, the training time of our algorithm is not significantly affected by the increase in the number of slave nodes.
In other words, if $m$ is large and $n$ is not that large, our algorithm training time will experience a rapid decrease as the number of slave nodes increases. This also indicates from another perspective that PIPADMM solving $\ell_1$ logistic regression is more efficient than existing parallel algorithms.

\begin{figure*}[h]\centering
	\includegraphics[width=14cm]{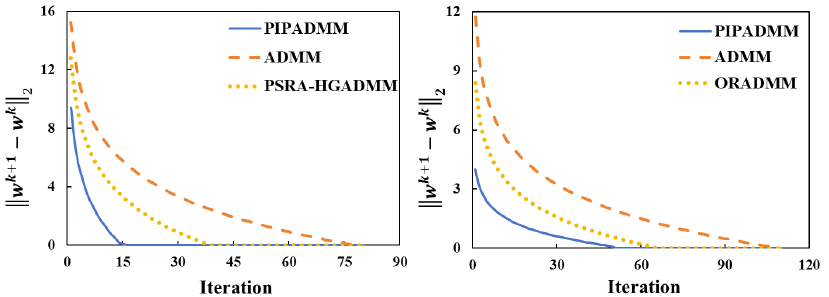}
	\caption{A schematic diagram showing the difference in the modulus of solutions between two adjacent iterations as the number of iterations increases (non parallel environment). On the left is $\ell_1$ logistic regression, and on the right is $\ell_1$ SVM.}
    \label{fig4}
\end{figure*}

In Table \ref{tab1},  “Test” is randomly selected 100 times for testing, and the mean is taken.  The results presented in Table  show that PIPADMM exhibits partition insensitivity when compared to the other two parallel algorithms. 
 This is reflected in the "Sparsity", "Train", and "Test" of the PIPADMM algorithm, which have the smallest changes as the number of slave nodes $D$ increases.
 Additionally, the findings in Table \ref{tab1} and Figure  \ref{fig4} highlight that PIPADMM demonstrates superior performance in terms of convergence speed, sparsity and prediction accuracy in the context of $\ell_1$ logistic regression calculation.

\begin{figure*}[h]\centering
	\includegraphics[width=14cm]{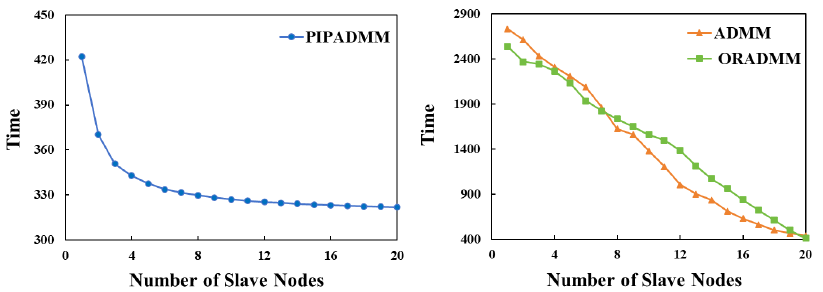}
	\caption{Illustration of the variation in training time for $\ell_1$ SVM with three parallel algorithms as the number of slave nodes changes.}
    \label{fig5}
\end{figure*}

\begin{table*}[!ht]\small
    \centering
    \renewcommand{\arraystretch}{1.2}
    \caption{Comparative analysis of Sparsity, training, and average testing accuracies (\%) for $\ell_1$ SVM using PIPADMM, ADMM, and ORADMM}
    \begin{tabular}{llllllllll}
    \hline
          & \multicolumn{3}{c}{PIPADMM} & \multicolumn{3}{c}{ADMM} & \multicolumn{3}{c}{ORADMM}\\ 
        \cmidrule(lr){2-4}\cmidrule(lr){5-7}\cmidrule(lr){8-10}
        D & Sparsity & Train & Test & Sparsity & Train & Test & Sparsity & Train & Test \\ \hline
        2 & 90.36 & 99.42 & 97.25 & 83.63 & 93.37 & 91.99 & 85.98 & 95.12 & 91.98 \\ 
        4 & 90.36 & 99.23 & 97.15 & 81.36 & 92.25 & 91.04 & 84.60 & 94.55 & 91.02 \\ 
        6 & 90.36 & 99.15 & 97.05 & 80.13 & 91.76 & 90.58 & 83.59 & 93.87 & 90.46 \\ 
        8 & 90.31 & 99.02 & 96.91 & 79.36 & 91.04 & 89.91 & 82.10 & 93.03 & 89.93 \\ 
        10 & 90.12 & 98.89 & 96.82  & 77.98 & 90.76 & 89.07 & 81.36 & 92.56 & 89.07 \\ 
        12 & 90.01 & 98.77 & 96.69 & 77.12 & 90.01 & 88.36 & 80.75 & 91.87 & 88.38 \\ 
        14 & 89.85 & 98.65 & 96.57 & 76.42 & 89.72 & 87.82 & 79.36 & 91.15 & 87.64 \\ 
        16 & 89.58 & 98.51 & 96.41 & 73.95 & 88.36 & 87.05 & 78.69 & 90.36 & 86.39 \\ 
        18 & 89.65 & 98.43 & 96.36 & 72.65 & 87.77 & 86.23 & 77.65 & 89.78 & 85.27 \\ 
        20 & 89.55 & 98.32 & 96.27 & 71.45 & 87.65 & 85.16 & 76.25 & 88.62 & 84.33 \\ \hline
    \end{tabular}
    \label{tab2}
\end{table*}

\subsubsection{Regularized support vector machine}\label{sec6.2}
Here, we compare the  parallel ADMM algorithms in  \cite{Boyd2010Distributed} (ADMM) and \cite{DAI2023Over} (ORADMM) with our algorithm (PIPADMM) for calculating $\ell_1$ SVM in \cite{Zhu2003l1}. 
Figure \ref{fig4} illustrates the convergence speed of three algorithms for solving $\ell_1$ SVM, with PIPADMM being the best and ORADMM being the second. The reason why PIPADMM is optimal is the main highlight of this paper, as our parallel ADMM algorithm requires the least number of variables per iteration. The faster performance of ORADMM compared to ADMM is attributed to its utilization of the over-relaxation acceleration technique described in \cite{Qin2016Recent}.

Figure \ref{fig5} describes the changes in training time of the three algorithm algorithms as the number of nodes increases.
Similar to  solving $\ell_1$ logistic regression, the training time of PIPADMM is less affected by the increase in the number of slave nodes.  The reason for this is also similar. The subproblem of $\bm z_d$ in the  $\ell_1$ SVM in (\ref{cz}) solving process is greatly affected by the number of rows in $\bm A_d$, while the update of  $\bm r_d$ in PIPADMM has an explicit solution, which can avoid this the computational burden.

The results in Table \ref{tab2} indicate that PIPADM has better sparsity and prediction performance in solving $\ell_1$ SVM compared to the other two parallel algorithms. Furthermore, in terms of predictive performance, $\ell_1$  SVM performs better than $\ell_1$  logistic regression in rcv1.binary datasets. This may be attributed to the fact that this dataset is linearly separable.

\section{Conclusion and Discussion}\label{sec7}
This paper considers a novel parallel ADMM algorithm framework to achieve distributed  model fitting. This framework greatly reduces the number of iteration variables required in ADMM iterations as it does not require transforming optimization problems into consensus problems. The fewer variables greatly reduce the computational and memory burden of parallel algorithms, thus allowing them to effectively handle high-dimensional datasets. In addition, this parallel framework also has a sample partitioning insensitivity, that is, the algorithm's solution is minimally affected by the sample parallelization method.

Although our algorithm has linear convergence when $\mathcal L$ and $\mathcal{R}$ are both convex, the convergence of the algorithm is challenging when $\mathcal L$ or/and $\mathcal{R}$ are non convex. 
Some studies have partially addressed this issue, such as  $\mathcal L$ requiring  to have first-order Lipschitz continuity (\cite{Guo2016Convergence}). However, there is still a lack of research on the generalization of  $\mathcal L$. Moreover, ADMM algorithm can be accelerated by some techniques, such as adaptive $\mu$ in \cite{Boyd2010Distributed} and \cite{Lin2022Alternating}, and over-relaxation in \cite{Qin2016Recent}. Our future work will focus on these two points to improve the efficiency and applicability of the proposed new parallel algorithm framework.
necessary.

%\section*{Acknowledgments}

\bibliographystyle{IEEEtran}
\bibliography{IEEEabrv, A_Partition-insensitive_Parallel_Framework_for_Distributed_Model_Fitting}


{\appendix[Supplementary proof of regression subproblem]\label{C}
 Here we supplement the closed-form solution of the proximal operator for the loss function of SVR. The loss function of SVR is called the $\varepsilon$-insensitive loss function, defined as follows
\begin{equation}\label{svr}
    L_{\varepsilon}( r_i) = \begin{cases}
        0 & \text{if } |r_i| \leq \varepsilon, \\
        |r_i| - \varepsilon & \text{otherwise}.
    \end{cases}
\end{equation} 
where  $\varepsilon$ is a specified tolerance value.

The proximal operator of $\varepsilon$-insensitive loss  is defined as
\begin{align}\label{svr2}
\mathop {\arg \min }\limits_{{r}_i}  L_{\varepsilon}( r_i) + \frac{\mu }{2}\left\| {{r}_i -{r}_i^0} \right\|_2^2
\end{align}
with $\mu > 0$ and a given constant ${r}_i^0$, has a closed-form solution defined as
\begin{small}
\begin{align}
r_i^*=
\begin{cases}
r_i^0, & \text{if } |r_i^0| < \varepsilon + \frac{1}{\mu}, \\
\text{sign}({r}_i^0) \varepsilon, & \text{if } |{r}_i^0| \in [\varepsilon, \varepsilon + \frac{1}{\mu}], \\
\text{sign}(r_i^0)\max\{0, |r_i^0| - \frac{1}{\mu} \}, & \text{if } |r_i^0| > \varepsilon + \frac{1}{\mu}.
\end{cases}
\end{align}
\end{small}

\begin{IEEEproof}
(1) If $|r_i| < \varepsilon$, the objective function in (\ref{svr2}) is:
\begin{equation}\label{proof1}
F(r) = \frac{\mu}{2}\left\| {r}_i - {r}_i^0 \right\|_2^2.
\end{equation}
Then, we have:
\begin{equation}
r_i^* = {r}_i^0.
\end{equation}
Please note that in this case, it is necessary to have $|{r}_i^0| < \varepsilon$.

(2) If $|r_i| > \varepsilon$, the objective function is:
\begin{equation}\label{proof2}
F(r) = (|r_i| - \varepsilon) + \frac{\mu}{2}\left\| {r}_i - {r}_i^0 \right\|_2^2.
\end{equation}
According to (\ref{lasso}), we have:
\begin{equation}\label{svr3}
{r}_i^* = \text{sign}({r}^0_i) \cdot \text{max}\{0, \left| {r}^0_i \right| - \frac{1}{\mu} \}.
\end{equation}
In this case, it is necessary to have $|{r}_i^0| > \varepsilon + \frac{1}{\mu}$.

Moving forward, our primary focus will be on identifying the optimal solution within the interval $|{r}_i^0| \in [\varepsilon, \varepsilon + \frac{1}{\mu}]$. It's worth noting that the objective functions in equations (\ref{proof1}) and (\ref{proof2}) share a quadratic form. Consequently, the minimization problem defined in equation (\ref{svr2}) only involves these two objective functions. Therefore, the optimal solution within the interval of interest must be either $\text{sign}({r}_i^0) \varepsilon$ or $\text{sign}({r}_i^0) (\varepsilon + \frac{1}{\mu})$. In the following discussion, we will only discuss the case where ${r}_i^0 > 0$, and the rest of the cases are similar.

To determine the optimal solution, we need to compare the values of the objective functions at these two points. It is clear that $F(\varepsilon) = \frac{\mu}{2}\left\|\varepsilon - {r}_i^0\right\|_2^2$, and $F(\varepsilon + \frac{1}{\mu}) = \frac{1}{\mu} + \frac{\mu}{2}\left\|\varepsilon + \frac{1}{\mu} - {r}_i^0\right\|_2^2$. Since it follows that $\frac{1}{\mu} > 0$ and $\varepsilon - {r}_i^0 \in [-\frac{1}{\mu}, 0]$, we have $F(\varepsilon + \frac{1}{\mu}) > F(\varepsilon)$. Therefore, if ${r}_i^0 \in [\varepsilon, \varepsilon + \frac{1}{\mu}]$, the optimal solution is ${r}_i^* = \varepsilon$.

Based on the above discussion, we arrive at the following expression for ${r}_i^*$,
\begin{small}
\begin{align}
{r}_i^* =
\begin{cases}
{r}_i^0, & \text{if } |{r}_i^0| < \varepsilon + \frac{1}{\mu}, \\
\text{sign}({r}_i^0) \varepsilon, & \text{if } |{r}_i^0| \in [\varepsilon, \varepsilon + \frac{1}{\mu}], \\
\text{sign}({r}_i^0)\max\{0, |{r}_i^0| - \frac{1}{\mu} \}, & \text{if } |{r}_i^0| > \varepsilon + \frac{1}{\mu}.
\end{cases}
\end{align}
\end{small}
\end{IEEEproof}}

\vfill

\end{document}